\documentclass{article}

\usepackage{PRIMEarxiv}

\usepackage[utf8]{inputenc} 
\usepackage[T1]{fontenc}    
\usepackage{hyperref}       
\usepackage{url}            
\usepackage{booktabs}       
\usepackage{amsfonts}       
\usepackage{nicefrac}       
\usepackage{microtype}      
\usepackage{lipsum}
\usepackage{fancyhdr}       
\usepackage{graphicx}       
\graphicspath{{media/}}     

\title{Prob-cGAN: A Probabilistic Conditional Generative Adversarial Network for LSD1 Inhibitor Activity Prediction
\thanks{\textit{\underline{Citation}}: 
\textbf{Hanyang Wang}} 
}

\author{
  Hanyang Wang
  \texttt{\{Hanyang Wang\}hanyang.wang2024@gmail.com} \\
}

\begin{document}
\maketitle

\begin{abstract} The inhibition of Lysine-Specific Histone Demethylase 1 (LSD1) is a promising strategy for cancer treatment and targeting epigenetic mechanisms. This paper introduces a Probabilistic Conditional Generative Adversarial Network (Prob-cGAN), designed to predict the activity of LSD1 inhibitors. The Prob-cGAN was evaluated against state-of-the-art models using the ChEMBL database, demonstrating superior performance. Specifically, it achieved a top-1 $R^2$ of 0.739, significantly outperforming the Smiles-Transformer model at 0.591 and the baseline cGAN at 0.488. Furthermore, it recorded a lower $RMSE$ of 0.562, compared to 0.708 and 0.791 for the Smiles-Transformer and cGAN models respectively. These results highlight the potential of Prob-cGAN to enhance drug design and advance our understanding of complex biological systems through machine learning and bioinformatics. \end{abstract}


\section{Introduction}
Epigenetic modifications such as histone methylation have been extensively studied for their crucial role in the regulation of genome-dependent biological processes \cite{huang2014snapshot}. Recent research has revealed the reversible nature of histone methylation, which is mediated by enzymes such as lysine-specific demethylase 1 (LSD1/KDM1A)\cite{shi2004histone}. LSD1 is a FAD-dependent oxidase that can demethylate different types of methylation while exhibiting functions as a transcription inhibitor and activator. The diverse functionalization of LSD1 has been linked to the dysregulation of gene expression and various pathogenic conditions such as tumorigenesis \cite{wang2007opposing}. Inhibition of LSD1 has emerged as a promising therapeutic strategy for cancer treatment and epigenetic targeting \cite{bansal2016emerging}.

Recent advances in the study of LSD1 and its inhibitors have led to the discovery of several effective compounds, which are currently undergoing clinical evaluation \cite{zhu2019lsd1}. These discoveries have been supported by the identification of structural and sequence similarities between LSD1 and monoamine oxidases (MAOs), suggesting that MAO inhibitors can function similarly as LSD1 inhibitors \cite{lee2006functional}. Since LSD1 can remove the transcriptional activating marks, it has the potential of aberrantly silencing the tumor suppressor genes, which can be reactivated by the inhibition of LSD1 and increasing methylation levels  \cite{murray2014re,huang2007inhibition}.

The complexity of biological systems and the vast amount of data generated in the field of drug discovery require advanced machine learning methods to facilitate the design of new drug candidates and the identification of potential targets. The study of LSD1 and its inhibitors 
offers insights into the relationships between structure, function, and biological activity, and may ultimately lead to the development of more effective treatments for a range of diseases. 
Several computational methods \cite{perkins2003quantitative, rastelli2002discovery, xu2019investigating, wang2019investigating} have investigated LSD1 inhibitors for predicting binding affinity and ligand-receptor interactions, as well as identifying key chemical groups in effective molecules for designing new drugs. For example, Xu et al. utilized a 3D-QSAR model to explore the binding modes of stilbene derivatives as LSD1 inhibitors \cite{xu2019investigating}. Wang et al. employed CoMFA and CoMSIA to generate a molecular model for (4-cyanophenyl)glycine derivatives \cite{wang2019investigating}. These analyses provide valuable information on the structure-activity relationship of LSD1 inhibitors. 
Recently, machine learning techniques have emerged to address the challenges of QSPR/QSAR modeling. For instance, an inductive transfer learning model, MolPMoFiT \cite{li2020inductive}, utilizes a molecular prediction architecture for QSPR/QSAR modeling, which is pre-trained on 1 million unlabeled molecules from the ChEMBL database via self-supervised learning and fine-tuned on smaller datasets with specific endpoints. Sakai et al. employed GCN models constructed from 2D structural data to accurately predict the activity of 127 diverse targets in the ChEMBL database \cite{sakai2021prediction}. These research works provide inspiration for improving QSPR/QSAR modelling using machine learning techniques.  

\begin{figure*}[htpb]
\centering
  \includegraphics[width=0.90\textwidth]{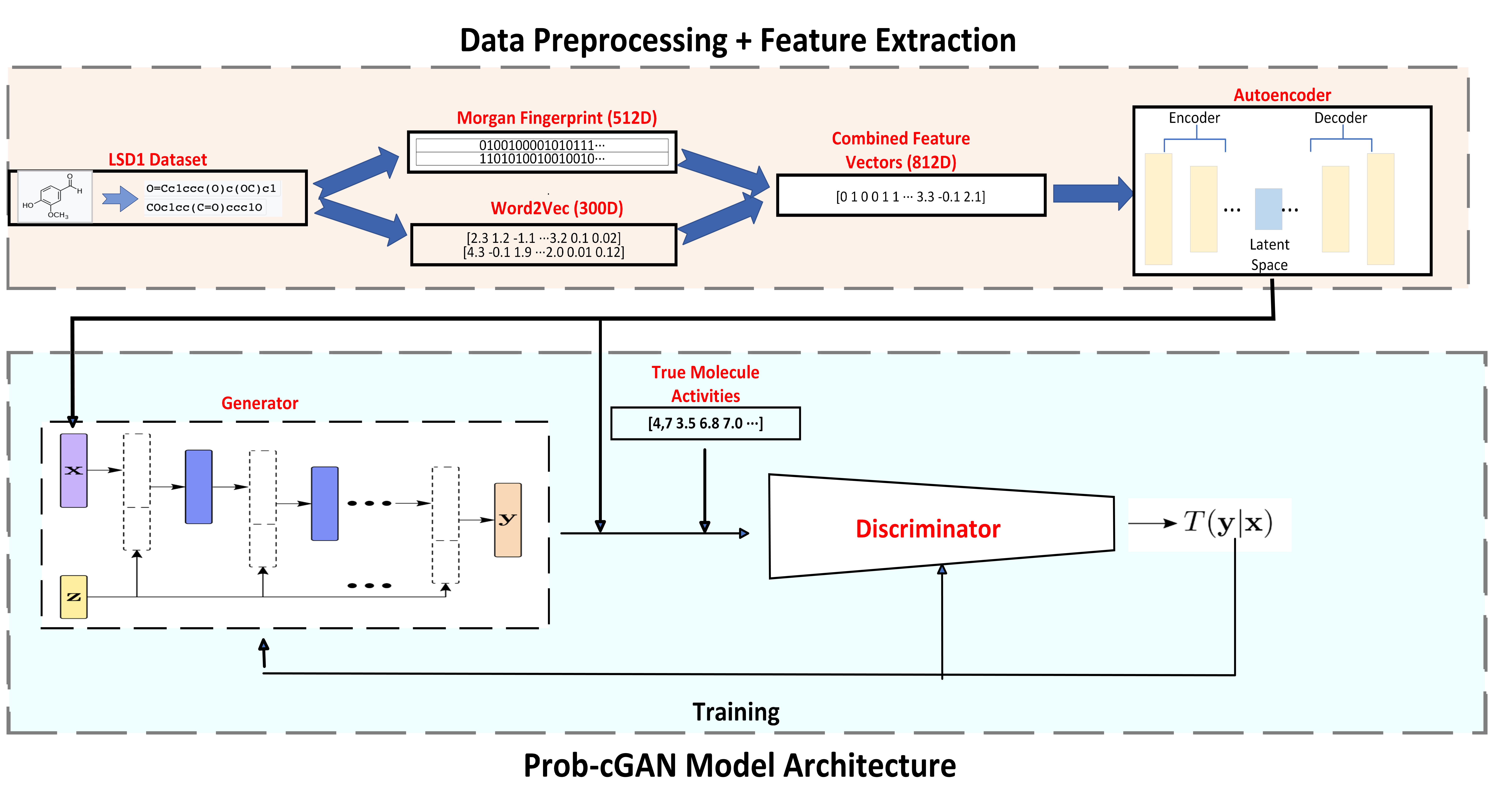}
\caption{Overview of the proposed Probabilistic Conditional Generative Adversarial Networks (CGAN) with molecule activity prediction. (1) SMILES strings of each molecule are converted into Morgan Fingerprint \cite{sandfort2020structure} and word embedding vector \cite{levy2014neural} descriptors, which are combined to form 812-dimensional vectors. (2) An autoencoder is applied to remove noisy features and extract a 203-dimensional latent space $x$. (3) The noise-injection generator incorporates the noise vector $z$ by concatenating it with the hidden representation at each layer, introducing randomness and variability into the generated molecules. (4) The discriminator receives input from the generated fake molecule activities, latent space vectors, and true activities, and produces a final one-dimensional output without any activation function, corresponding to "T" in the f-GAN framework.}
  \label{general-description}
\end{figure*}

Existing machine learning methods for anticipating chemical reactions often face some challenges. 
1) Two types of molecule descriptors have been widely used, ie, Morgan Fingerprint \cite{sandfort2020structure}, and SMILES strings \cite{o2012towards}. Morgan Fingerprint encodes the 2D molecular structure through a bit sequence, while SMILES strings offer a unique textual representation of the molecular formula.  These two descriptors describe a molecule from different aspects. It is hence advantageous to combine these two descriptors for better accessing the characteristics of the molecules. 
2) Previous machine learning methods for drug discovery, eg, Decision Tree \cite{song2015decision}, Support Vector Regression (SVR) \cite{noble2006support} and XGBoost \cite{chen2015xgboost} often assume that the test data distribution follows closely the training data distribution, while ignoring the distribution discrepancy between these two. This may cause the model built on the basis of the training data to perform poorly on the novel test data, leading to overfitting problems. 
This can result in models that are not robust to changes in the underlying data distribution. 

In this paper, both Morgan fingerprint and SMILES strings are utilized to describe the molecule structure, and a Probabilistic Conditional Generative Adversarial Network (Prob-cGAN) is proposed to accurately model the complex data distributions for LSD1 inhibitor activity prediction. The proposed Prob-cGAN works well even when the probability density functions are unknown \cite{oskarsson2020probabilistic}, which makes it particularly useful for predicting the activity of small molecules with highly variable or unknown activity distributions. 
Moreover, Prob-cGAN can capture important data properties without requiring any extensions and can handle high-dimensional features and target variables well.
Given its ability to model complex conditional distributions effectively, Prob-cGAN has great potential as a tool for probabilistic activity regression tasks. 
Compared to previous research work using the Conditional Generative Adversarial Network (cGAN) as a regressor
\cite{aggarwal2019benchmarking}, the proposed Prob-cGAN is unique as it uses a probabilistic model to predict activity in small molecules for the first time, which could better model complex conditional distributions than traditional cGAN regressors \cite{aggarwal2019benchmarking}. 

\section{Feature descriptor} The data processing pipeline of the model comprises two main parts. The first part involves the Morgan Fingerprint and Word Embedding, utilized for feature extraction. The second part employs an autoencoder model to reduce the dimensions of features and eliminate excessive redundancy in the merged data. Subsequently, the Prob-cGAN model leverages these processed features to predict molecular activities.

Structure-based descriptors, such as Morgan fingerprints, are widely used to encode structural information and have been applied across a broad range of applications to predict molecular activities~\cite{sandfort2020structure}. In this study, Morgan fingerprints are generated for each SMILES string in the dataset using the RDKit package (version 2020.09.1), with a preset length of $L = 512$ and a radius $r = 3$. To enhance the discriminative power of the features, Word2Vec~\cite{church2017word2vec} is employed to generate word embeddings for SMILES strings. Specifically, a pre-trained model is used to embed SMILES strings as 300-dimensional vectors in high-dimensional space. The inclusion of word embedding of SMILES strings significantly improves the prediction accuracy, as demonstrated later.

However, the combination of Morgan fingerprints and word embeddings results in redundant features and high dimensionality, which can negatively affect the prediction performance of the model. To address this issue, Bengio et al. \cite{bengio2006greedy} suggested the use of an autoencoder designed to preserve the latent data structure and derive a compact feature representation. This approach effectively reduces the original 812-dimensional features to 203-dimensional features.

\section{Prob-cGAN}
The proposed Prob-cGAN produces an accurate prediction of LSD1 inhibitor activity based on the derived features. Due to the uncertainties in the molecule structure information, and noise which is from the traditional text feature extraction methods of LSD1 inhibitors, the traditional black-box machine learning regression algorithms\cite{xu2019review} derived from the training data often do not work well on the novel testing data. 
In contrast, the proposed Prob-cGAN could  by learning the underlying data distribution and generating new samples, effectively handle the uncertainties and noise present. Further Prob-cGAN's ability to generate probabilistic predictions allows for a more comprehensive assessment of the inhibitor activity. Instead of providing a single point estimate, Prob-cGAN can estimate the probability distribution of the activity, providing a measure of uncertainty associated with each prediction. This feature is particularly valuable when dealing with novel testing data, as it helps account for the distribution discrepancy between the training and testing datasets.

More specifically, the proposed Prob-cGAN consists of a generator network and a discriminator network. 
The generator model utilizes a deep neural network regression for predicting drug efficacy by combining preprocessed molecule structured data and random noise. It uses a noise injection architecture that enables the generator to learn the optimal position for noise injection in the network, allowing it to accommodate different types of noise and distributional assumptions. The adaptability of the generator to accommodate diverse noise types and distributional assumptions further enhances its utility, ensuring that it remains robust and effective across various scenarios and datasets. The generator's crucial function is to derive accurate predictions of drug efficacy by synthesizing potential molecular structures based on the provided data. Notably, this synthesis is not a simple replication but an innovative creation of molecular structures, showcasing the generator's ability to generate novel data that aligns with desired drug efficacy outcomes. This feature is particularly significant for a research paper, as it demonstrates the model's potential to drive drug discovery and development by presenting a diverse set of potential drug candidates for exploration.

The discriminator network within the Prob-cGAN model plays a crucial role, employing an f-GAN loss function to assess the divergence between two probability distributions. This loss function is meticulously tailored to suit the specific needs and characteristics pertinent to drug discovery. The discriminator’s main function is to meticulously evaluate and differentiate between the input molecular data and the data generated by the generator network.

One of the defining features of the discriminator is its dual-pathway architecture. It comprises separate pathways for handling input and generated data, allowing the network to concentrate on and highlight the unique molecular features present in each data stream. This dual-pathway design is not just a structural novelty but serves a vital purpose in the model’s functionality.

By providing a dedicated pathway for each data type, the discriminator can conduct a more precise analysis of the molecular characteristics. This in-depth scrutiny is essential for verifying the authenticity of the data and ensuring its alignment with the targeted drug efficacy outcomes.

The specific distinction between genuine molecular traits and synthesized attributes plays a pivotal role in refining the generator’s output. By doing so, the discriminator directly contributes to the enhancement of the model’s overall predictive accuracy. The end result is a more reliable and effective model, capable of generating high-quality predictions that are invaluable in the realm of drug discovery.
The proposed Prob-cGAN offers a novel way to handle the LSD1 inhibitor activity prediction, which often lacks a large-scale dataset and contains large variations in datasets.

\section{Dataset description}
Our LSD1 dataset is obtained from the ChEMBL database (version 28) \cite{gaulton2017chembl}, which consists of 931 compounds that target LSD1. The dataset contains the molecule structures in terms of SMILES strings, the corresponding biological activity, 
To ensure that the dataset is suitable for model training, we extracted assay descriptions and documents from ChEMBL and filtered out biological activities that were similar but did not display the characteristics of LSD1 inhibition. 
and the pChEMBL values used to inhibit LSD1, which serve as the regression target for the model.

The dataset is randomly divided into two subsets: 80\% for training and 20\% for testing. The training dataset is utilized to develop the model, while the testing dataset is employed to evaluate the model's performance.

\subsection{Evaluation Metrics} Two evaluation metrics are utilized to assess the prediction accuracy of the models, namely Root Mean Square Error (RMSE) and $R^2$. RMSE quantifies the difference between the predicted values and the actual values \cite{willmott2005advantages}, serving as a measure of prediction error. Conversely, $R^2$ assesses the proportion of variance in the dependent variable that can be explained by the independent variables in the model \cite{draper1998applied}. A lower RMSE or a higher $R^2$ score indicates a more accurate model. Both RMSE and $R^2$ are widely used to compare the performance of different models \cite{hawkins2003assessing}.

\begin{table}[h]
\caption{Comparison against different methods}
\centering
\begin{tabular}{|l|c|c|}
\hline
\textbf{Methods} & \textbf{$R^2$} & \textbf{$RMSE$} \\
\hline
SVR                               & 0.6642 & 0.6484 \\
\hline
Ridge                             & 0.6428 & 0.6688 \\
\hline
Random Forest                     & 0.6686 & 0.6441 \\
\hline
Decision Tree                     & 0.4521 & 0.8283 \\
\hline
XGBoost                           & 0.6401 & 0.6713 \\
\hline
MLP                               & 0.5465 & 0.7535 \\
\hline
K-Neighbours                      & 0.5816 & 0.7237 \\
\hline
Smiles-Transformer                & 0.6097 (±0.0078) & 0.6990 (±0.0070) \\
\hline
\textbf{Prob-cGAN}                & \textbf{0.8131 (±0.0008)} & \textbf{0.4838 (±0.0010)} \\
\hline
\end{tabular}
\end{table}

\begin{figure*}[htpb]
\centering
  \includegraphics[width=0.90\textwidth]{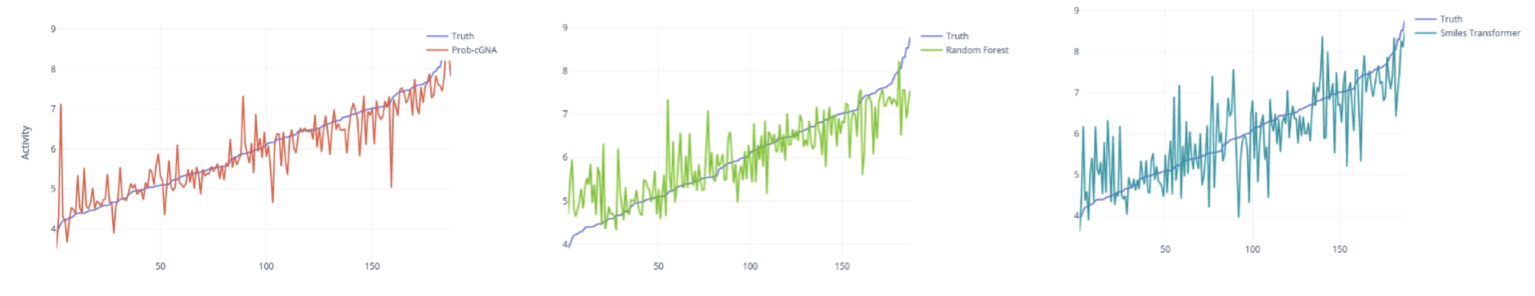}
  \caption{Model results of activity prediction.
  }
  \label{fig:model-results}
\end{figure*}

\begin{figure*}[htpb]
\centering
   \includegraphics[width=0.8\linewidth]{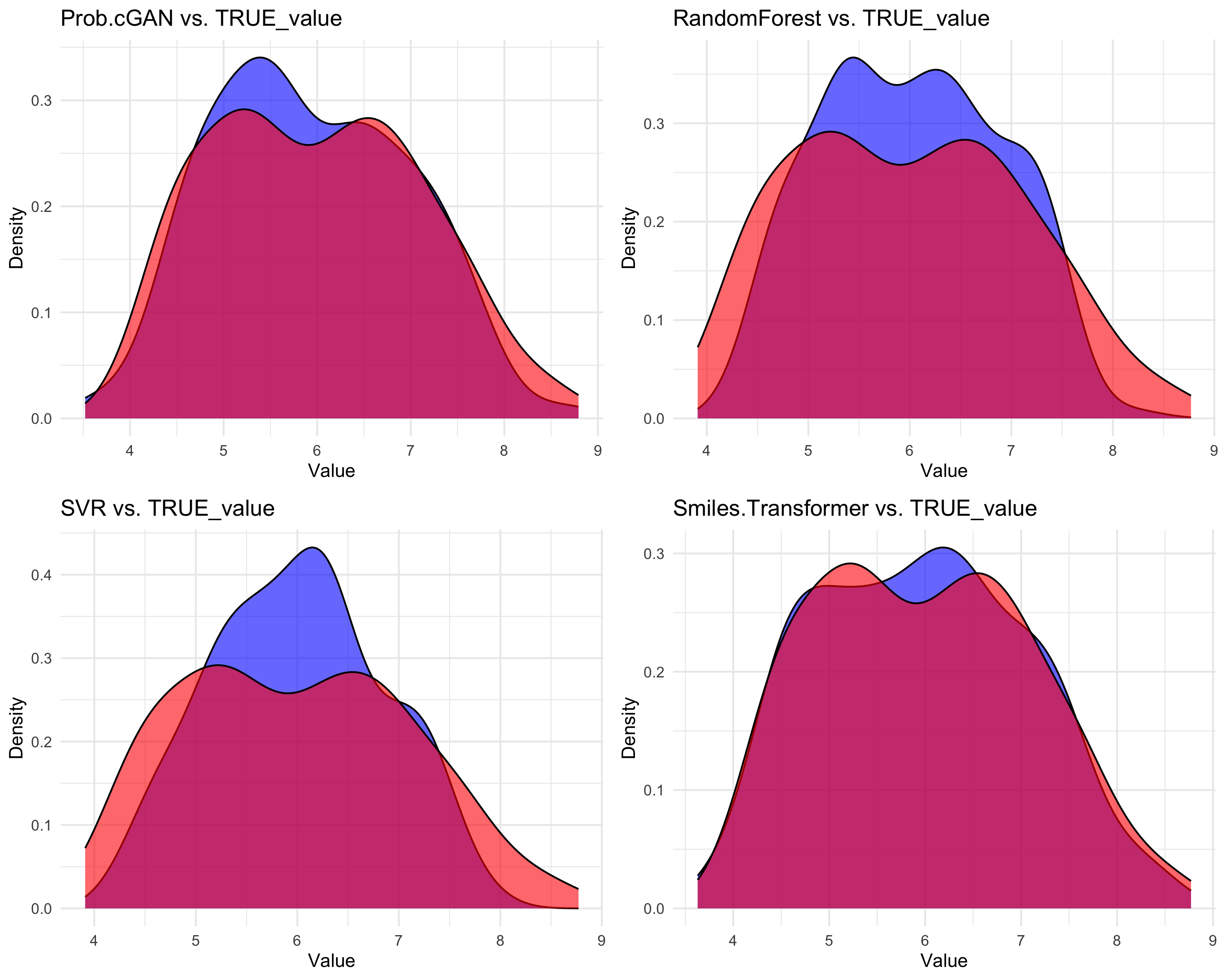}
   \caption{Model results of activity prediction distribution}
   \label{fig:model-results}
\end{figure*}

\section{Result} 
\subsection{Result: Comparsion against Machine Learning Algorithms}
The proposed Prob-cGAN is compared with state-of-the-art models for the prediction of the activity of the LSD1 inhibitor. The results are summarized in Table~\ref{fig:model-results}. The following can be observed from Table~\ref{fig:model-results}. 1) The proposed Prob-cGAN significantly outperforms all the compared methods. Compared to the previously best performing method, Random Forest, the proposed Prob-cGAN largely increases $R^2$ from 0.6686 to 0.8131, and significantly reduces the RMSE from 0.6441 to 0.4838. 2)Previously, models such as SVR and Random Forest have shown their prowess in predicting LSD1 inhibitor activity, with Random Forest's set of decision trees providing a robust framework against variations in data distribution \cite{liaw2002classification}. However, the proposed Prob-cGAN model stands out by incorporating an advanced strategy to adeptly navigate uncertainties. Through its noise injection architecture within the generator network, it introduces adaptability to various noise types and distributional challenges, showcasing a significant edge in handling data inconsistencies \cite{goodfellow2014generative}. This approach to managing uncertainties translates into a remarkable enhancement in both robustness and accuracy, ensuring a substantial performance leap in predicting drug efficacy amidst uncertainties \cite{kingma2013auto}. 3) The proposed method is also compared with SMILES-transformer \cite{honda2019smiles}, which has achieved excellent performance in molecular property prediction tasks. However, due to the discrepancy in data distribution, the SMILES transformer struggles to accurately predict the activity of the LSD1 inhibitor. On the contrary, the proposed Prob-cGAN demonstrates a remarkable ability to adapt to these discrepancies, handling data uncertainties with its innovative noise injection architecture and probabilistic modeling approach. This enables Prob-cGAN to offer more accurate and robust predictions for LSD1 inhibitor activity, as evidenced by its significant improvement in performance metrics; it increases the value of $R^2$ from 0.6097 to 0.8131 and reduces the RMSE from 0.6990 to 0.4838.
In summary, Table~\ref{fig:model-results} shows that the Prob-cGAN model efficiently predicts molecular activity while adjusting for data uncertainty by incorporating multiple data sources and advanced modelling techniques. The proposed method has the potential of broader applications in inhibitor discovery and drug development, offering a promising solution for improving the accuracy and efficiency of predicting molecular activities. 

\subsection{Result: Comparison against Random Forest and Smile Transformers.}

We also assessed the Smiles-transformer model of Honda et al. \cite{honda2019smiles} and found that it outperformed other machine learning models, such as KNN and Decision Tree, on small-scale datasets. However, compared to SVR, Random Forest, Ridge, and XGBoost, it still had certain disadvantages. This demonstrates Smiles-transformer's relatively complete feature processing and prediction capabilities, allowing relatively accurate prediction of molecular properties. Moreover, incorporating the word-embedding Smiles vector into our feature engineering is reasonable.

Fig. 2 presents a visual comparison of the prediction results between the proposed prob-cGAN model and other prominent machine learning models, including the Smiles Transformer (ST). In the graph, molecular activities in the test set are sorted in ascending order to facilitate a straightforward comparison.

Smiles Transformer, praised for its capability in molecular property prediction \cite{honda2019smiles}, exhibits proficiency in grasping complex molecular structures. However, its predictions show noticeable fluctuations at certain activity levels, such as around 20, 60, and 160. This behavior may highlight its sensitivity to novel chemical properties that are not well represented in the training set, a challenge commonly faced in molecular activity prediction.

However, the prob-cGAN model displays a prediction curve that closely aligns with the true values, demonstrating exceptional consistency across varying activity levels. This robustness, even in the presence of chemically diverse molecules, is a testament to its advanced capability to handle uncertainty of data and variations in distribution, a crucial aspect in drug discovery applications \cite{chen2018rise}.

Performance-wise, the prob-cGAN model stands out with a $R^2$ value of 0.8131 and an RMSE of 0.4838, significantly outperforming the Smiles Transformer. This underscores the superior predictive accuracy of the model, making it a highly reliable choice to predict molecular activities, particularly in the discovery of inhibitors and drugs, where precise predictions are paramount \cite{gawehn2016deep}.

\subsection{Results: Prob-cGAN vs. cGAN.}

\begin{table}[htpb]
\caption{Ablation study.} 
\small
\centering
\begin{tabular}{|c|c|c|}
\hline
\textbf{Methods} & \textbf{$R^2$} & \textbf{$RMSE$} \\ \hline
\textbf{Prob-cGAN}                         
& \textbf{0.813} & \textbf{0.484} \\ \hline
cGAN & 0.488 & 0.791           \\ \hline
(w/0) Autoencoder & 0.413 & 0.870 \\ \hline
(w/0) Autoencoder, Word Embedding & 0.258 & 0.970 \\ \hline
\end{tabular}
\end{table}

In this ablation study, we primarily focus our attention toward contrasting the performance of the Prob-cGAN model with the conventional cGAN, aiming to underscore the distinctive enhancements achieved by our proposed model.

The cGAN model, recognized for its utility in various domains, has demonstrated significant effectiveness, particularly in image generation and data augmentation tasks \cite{goodfellow2014generative}. It has established itself as a reliable model for generating synthetic data that closely resembles real data distributions, contributing to advancements in machine learning and artificial intelligence.

However, our study unveils a different narrative when these models are applied to the task of predicting molecular activities. The results clearly show that Prob-cGAN, with an $R^2$ of 0.813 and an RMSE of 0.484, significantly outperforms cGAN, which scores an $R^2$ of 0.488 and an RMSE of 0.791. This disparity in performance is noteworthy, especially considering the widespread success of cGAN in other applications.

The exceptional performance of Prob-cGAN stems from its innovative architecture, which incorporates the f-GAN framework, and the application of autoencoder and word embedding features for advanced data pre-processing and feature extraction. The f-GAN framework, renowned for its efficiency in approximating divergences between probability distributions, plays a critical role in the model’s ability to accurately capture the intricacies of molecular activity data \cite{nowozin2016f}.

Autoencoders, by design, excel in reducing the dimensionality of data and capturing latent representations, improving the model's ability to discern underlying patterns in molecular descriptors \cite{hinton2006reducing}. Inclusion of word embeddings further fortifies the model feature set, providing rich semantic relationships between molecular components, a technique well established in natural language processing and increasingly applied in cheminformatics \cite{duvenaud2015convolutional}.

Together, these elements equip Prob-cGAN with a robust framework, adept at navigating the complexities of molecular data, managing uncertainties, and delivering predictions of molecular activity that are both precise and reliable. This results in a noticeable performance boost over conventional cGAN models, cementing Prob-cGAN’s role as a valuable asset in the domain of drug discovery and molecular activity prediction.

In essence, while the cGAN has undoubtedly proven its capabilities in various fields, it is Prob-cGAN that shines in the realm of molecular activity prediction. Demonstrating unparalleled performance, Prob-cGAN solidifies its potential as a vital tool in drug discovery and other related fields, potentially transforming future research and clinical applications through its integration and scalability.

\section{Conclusion}
In this research, we unveiled the Prob-cGAN, an innovative deep learning methodology tailored for molecular property predictions, and performed an exhaustive analysis in comparison to other prevalent machine learning models. The results of our experiments unequivocally confirm that Prob-cGAN stands out in terms of both precision and precision, highlighted by a substantial increase in $R^2$ and a significant decrease in $RMSE$.

The distinctive strength of Prob-cGAN stems from its capacity to probabilistically navigate through intricate data distributions, rendering it a powerful instrument for molecular property predictions across a myriad of chemical applications, ranging from drug discovery to material science.

Looking ahead, there is a wealth of opportunity for future investigations to extend the application of Prob-cGAN to larger and more diverse datasets, as well as to explore its potential in various other chemical domains. The scalability and adaptability of Prob-cGAN mark it as a promising candidate for broad adoption in diverse scientific and industrial endeavors, opening new horizons for innovation and discovery.

\section*{Acknowledgments}

\bibliographystyle{unsrt}  
\bibliography{references}

\end{document}